\documentclass[
    ,final            
  ,numberedheadings 
]  {aipproc}

\layoutstyle{6x9}

\begin{document}

\title{Effects of LatticeQCD EoS and Continuous Emission on 
 Some Observables}



\author{Y.~Hama}{
  address={Instituto de F\'{\i}sica, Universidade de S\~ao 
           Paulo, Brazil}
}
\author{R.~Andrade}{
  address={Instituto de F\'{\i}sica, Universidade de S\~ao 
           Paulo, Brazil}
}
\author{F.~Grassi}{
  address={Instituto de F\'{\i}sica, Universidade de S\~ao 
           Paulo, Brazil}
}
\author{O.~Socolowski}{
  address={Instituto Tecnol\'ogico da Aeron\'autica, Brazil}
}
\author{T.~Kodama}{
  address={Instituto de F\'{\i}sica, Universidade Federal do 
           Rio de Janeiro, Brazil}
}
\author{B.~Tavares}{
  address={Instituto de F\'{\i}sica, Universidade Federal do 
           Rio de Janeiro, Brazil}
}
\author{S.S.~Padula}{
  address={Instituto de F\'{\i}sica Te\'orica, Universidade 
           Estadual Paulista}
}

\begin{abstract}
Effects of lattice-QCD-inspired equations of state and 
continuous emission on some observables are discussed, by 
solving a 3D hydrodynamics. The particle multiplicity as well 
as $v_2$ are found to increase in the mid-rapidity. We also 
discuss the effects of the initial-condition fluctuations. 
\end{abstract}

\maketitle


\section{HYDRODYNAMIC MODELS}
\label{ingredients}

Hydrodynamics is one of the main tools for studying the 
collective flow in high-energy nuclear collisions. Here, we 
shall examine some of the main ingredients of such 
a description and see how likely more realistic treatment of 
these elements may affect some of the observable quantities. 
The main components of any hydrodynamic model are the initial 
conditions, the equations of motion, equations of state and 
some decoupling prescription. We shall discuss how these 
elements are chosen in our studies. 
\bigskip

\noindent {\bf Initial Conditions}: 
In usual hydrodynamic approach, one assumes some highly 
symmetric and smooth initial conditions (IC). However, since 
our systems are small, large event-by-event fluctuations are 
expected in real collisions, so this effect should be taken 
into account. We introduce such IC fluctuations by using an 
event simulator. As an example, we show here the energy density for central Au+Au collisions at 130A GeV,\hfilneg\ 
\vspace*{.1cm} 
\begin{figure}[h] 
\includegraphics[angle=270,width=7.cm]{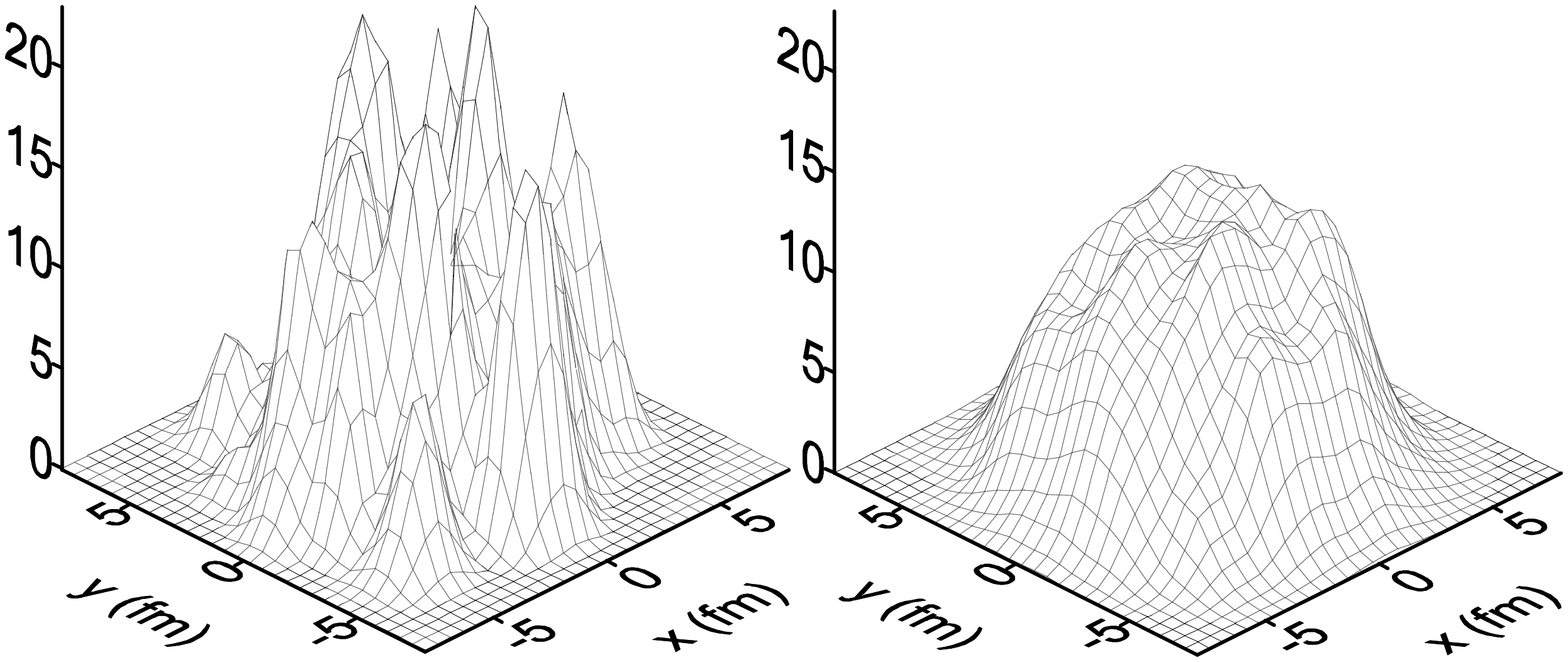}
\caption{The initial energy density at $\eta=0$ is plotted in 
 units of GeV/fm$^3$. One random event is shown vs. average 
 over 30 random events ($\simeq$ smooth initial 
 conditions in the usual hydro approach).}
\label{fig:IC}
\vspace*{-.5cm} 
\end{figure}

\noindent given by NeXuS\footnote{Many other simulators, based 
on microscopic models, {\it e.g.} HIJING \cite{hijing}, 
VNI \cite{vni}, URASiMA \cite{urasima}, $\cdots$, show such 
event-by-event fluctuations.} \cite{nexus}. Some consequences of such fluctuations have been 
discussed elsewhere\cite{fluctuations,hbt-prl,review}. We 
shall discuss some others in Sec.\ref{results}. 
\smallskip

\noindent {\bf Equations of Motion}: 
In hydrodynamics, the flow is governed by the continuity 
equations expressing the conservation of energy-momentum, 
baryon-number and other conserved charges. Here, for simplicity, we shall consider only the energy-momentum and 
the baryon number. Since our systems have no symmetry as discussed above, we developed a special 
numerical code called SPheRIO ({\bf S}moothed {\bf P}article 
{\bf h}ydrodynamic {\bf e}volution of {\bf R}elativistic heavy 
{\bf IO}n collisions) \cite{spherio}, based on the so called 
Smoothed-Paricle Hydrodynamics (SPH) algorithm \cite{sph}. 
The main characteristic of SPH is the parametrization of the 
flow in terms of discrete Lagrangian coordinates attached to 
small volumes (called ``particles'') with some conserved 
quantities. 
\smallskip

\noindent {\bf Equations of State}: 
In high-energy collisions, one often uses equations of state 
(EoS) with a first-order phase transition, connecting a 
high-temperature QGP phase with a low-temperature hadron phase. 
A detailed account of such EoS may be found, for instance, in 
\cite{review}. We shall denote them 1OPT EoS. However, lattice 
QCD showed that the transition line has a critical end point 
and for small net baryon surplus the transition is of crossover 
type~\cite{LQCD}. 
The following parametrization may reproduce this behavior, 
in practice:  
\vspace*{-.2cm} 
\begin{eqnarray}
  P  &=& \lambda P_H+(1-\lambda)P_Q 
        +2\delta/{\sqrt{(P_Q-P_H)^2+4\delta}}\ ,\\
  s  &=& \lambda s_H+(1-\lambda) s_Q\,, \\
  \epsilon &=& \lambda\epsilon_H+(1-\lambda)\epsilon_Q 
        -{2\,\left[1+(\mu/\mu_c)^2\right]\,\delta}/
         {\sqrt{(P_Q-P_H)^2+4\delta}}\ , 
\end{eqnarray}    
\vspace*{-1.cm} 
\begin{figure}[!h]
\hspace{4.cm}
\includegraphics[height=8.2cm,width=17.5cm]{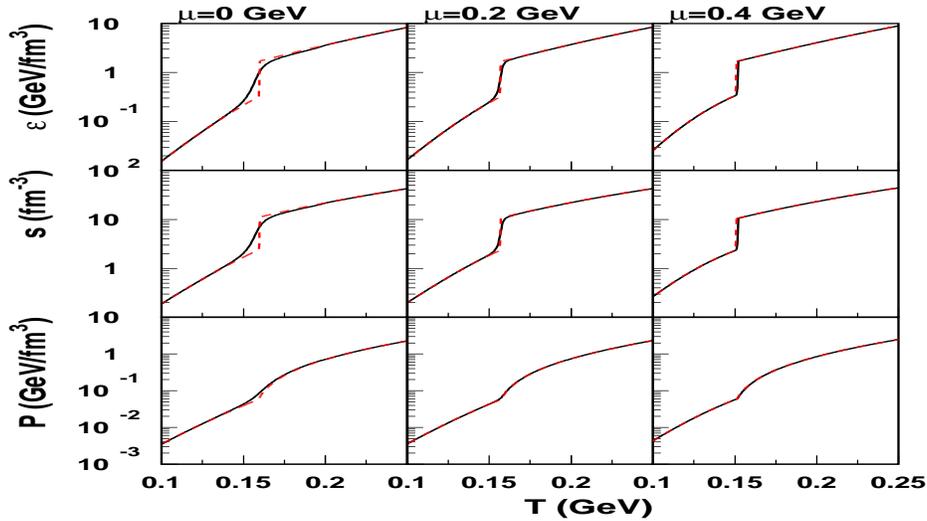} 
\caption{A comparison of $\varepsilon(T)$, $s(T)$ and $P(T)$ 
 as given by our parametrization with a critical point (solid 
 lines) and those with a first-order phase transition (dashed 
 lines).} 
\label{fig:EoS1}
\end{figure}

\noindent where 
$\lambda\equiv[1-(P_Q-P_H)/\sqrt{(P_Q-P_H)^2+4\delta}\ ]/2$ 
and suffixes $Q$ and $H$ denote those quantities given by 
the MIT bag model and the hadronic resonance gas, respectively, 
and $\delta\equiv\delta(\mu_b)=\delta_0\exp[-(\mu_b/\mu_c)^2]$, 
with $\mu_c=$const. 
As is seen, when $\delta(\mu_b)\not=0$, the transition from 
hadron phase to QGP is smooth. We could choose $\delta(\mu_b)$ 
so to make it exactly 0 when $\mu_b>\mu_c\,$, to guarantee the 
first-order phase transition there. However, in practice our choice above showed to be enough. We shall 
denote the EoS given above, with $\delta_0\not=0$, CP~EoS. 
Let us compare, in Figure~\ref{fig:EoS1}, $\varepsilon(T)$, $s(T)$ and $P(T)$, given by the two 
sets of EoS. one can see that the crossover 
behavior is correctly reproduced by our parametrization for 
CP EoS, while finite jumps in $\varepsilon$ and $s$ are 
exhibited by 1OPT EoS, at the transition temperature. It is 
also seen, as mentioned above, that at $\mu_b\sim0.4\,$GeV 
the two EoS are indistinguishable. Now, since in a 
real collision what is directly given is the energy distribution at a certain initial time (besides $n_b$, $s$, 
{\it etc.}), whereas $T$ is defined with the use of the former, 
we plotted some quantities as function of $\varepsilon$ in 
Figure \ref{fig:EoS2}. One immediately sees there some 
remarkable 
differences between the two sets of EoS: naturally $p$ is not 
constant for CP EoS in the crossover region; moreover, $s$ is 
larger. We will see in Sec.\ref{results} that these features 
affect the observables in non-negligible way.  
\bigskip
 
\noindent {\bf Decoupling Prescription}: 
Usually, one assumes decoupling on a sharply defined hypersurface. We call this {\it Sudden Freeze Out} (FO). 
However, since our systems are small, particles may escape 
from a layer with thickness comparable with the systems' sizes. 
We proposed an alternative description called 
{\it Continuous Emission} (CE)~\cite{ce} which, as compared to  FO, we believe closer to what happens in the actual collisions. 
In CE, particles escape from any space-time point $x^\mu$, 
according to a momentum-dependent\hfilneg\ 
\begin{figure}[!t]
\hspace{4.cm}  
\includegraphics*[height=7.cm,width=17.5cm]{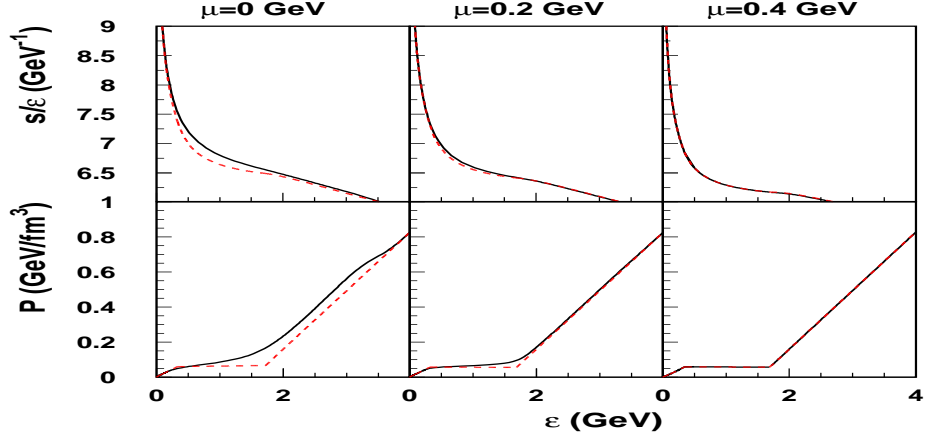} 
\caption{Plots of $s/\varepsilon$ and $P$ as function of 
 $\varepsilon$ for the two EoS shown in Figure~\ref{fig:EoS1}.}
\vspace*{-1.6cm}
\label{fig:EoS2}
\end{figure}


\noindent escaping probability 
${\cal P}(x,k)
=\exp\left[-\int_\tau^\infty \rho(x^\prime)\,\sigma v\; \mathrm{d}\tau^\prime\right].$ 
To implement CE in SPheRIO code, we had to approximate it to make the computation practicable. We 
took ${\cal P}$ on the average, {\it i.e.}, 
\begin{equation} 
  {\cal P}(x,k)\rightarrow\langle{\cal P}(x,k)\rangle 
   \equiv {\cal P}(x)
   =\exp\left(-\kappa\ s^2/|\mathrm{d}s/\mathrm{d}\tau|\right). 
\label{eq:prob}  
\end{equation} 
The last equality has been obtained by making a linear 
approximation of the density 
$\rho(x^\prime)=\alpha\, s(x^\prime)$ and 
$\kappa = 0.5\,\alpha\,\langle\sigma v\rangle$ is 
estimated to be 0.3$\,$, corresponding to 
$\langle\sigma v\rangle\approx$ 2~fm$^2$. It will be shown in 
Sec. \ref{results} that CE gives important changes in some 
observables.  
 
\section{RESULTS}
\label{results} 

Let us now show results of computation of some observables, as 
described above, for Au+Au  at 200A GeV. We start  
computing $\eta$ and $p_T$   
distributions for charged particles, to fix the parameters. 
Then, $v_2$ and HBT radii are computed free of parameters. 
\begin{figure}[!t]
\vspace*{-1.4cm}
\includegraphics*[scale=0.22]{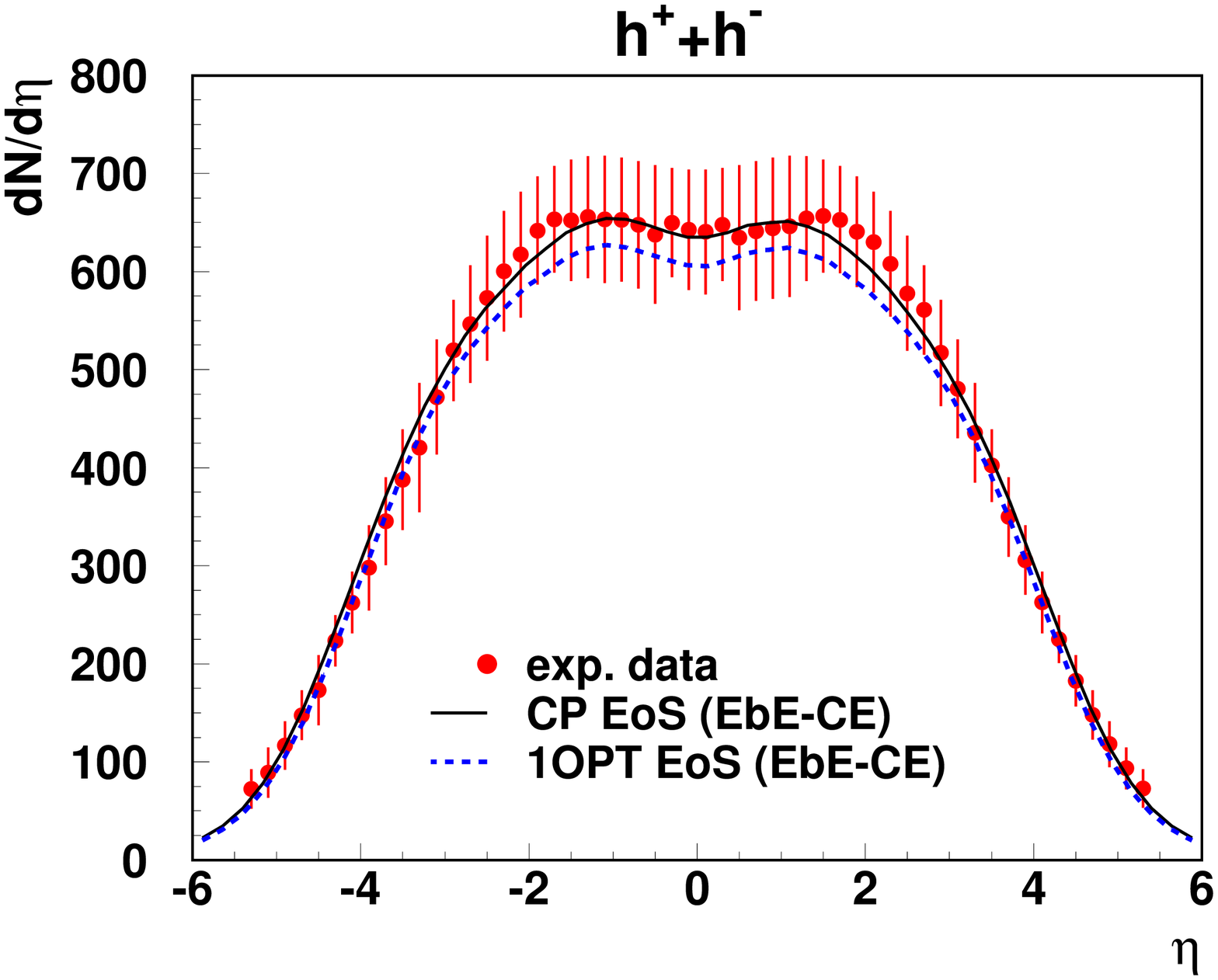}
\includegraphics*[scale=0.22]{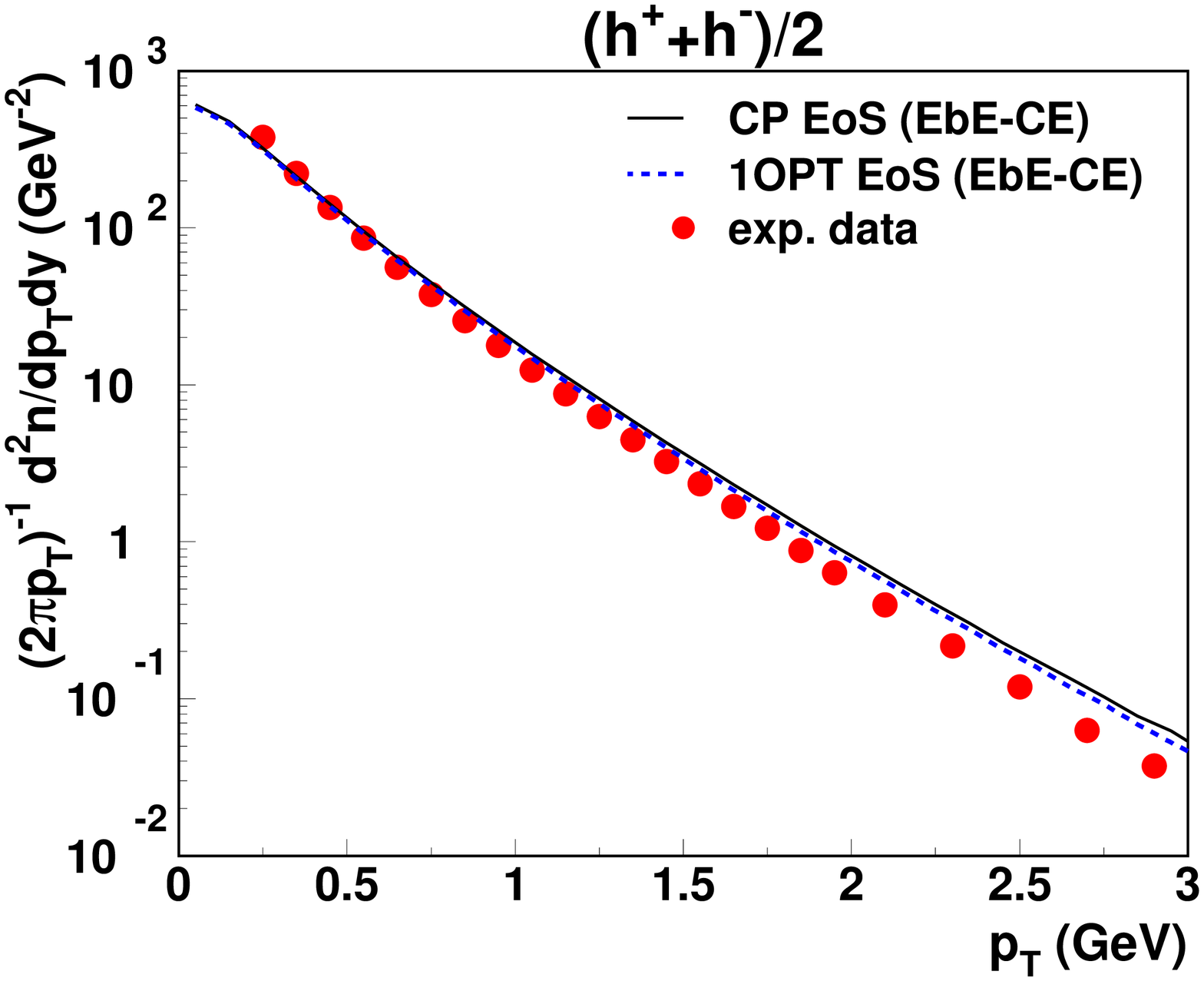}
\vspace*{-2.cm} 
\caption{$\eta$ and $p_T$ distributions for the most central 
 Au+Au at 200A GeV. Results of CP EoS and 1OPT EoS are 
 compared. The data are from PHOBOS Collab.\cite{phobos1}.}
\label{fig:eta}
\vspace*{-1.5cm} 
\end{figure}
\begin{figure}[!b]
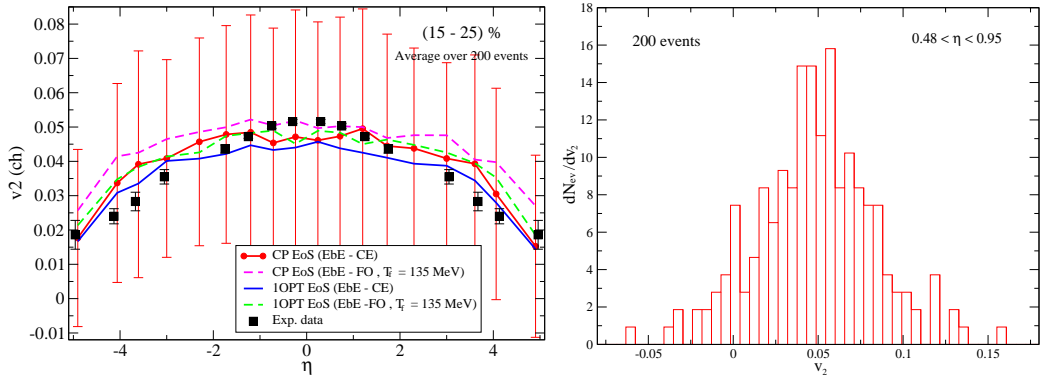

\includegraphics*[scale=0.3]{v2.eps} 
\vspace*{-.5cm}
\vspace*{-1.2cm}
\includegraphics*[scale=0.29]{dNdv2.eps} 
\vspace*{-.5cm}
\caption{Left: $\eta$ distribution of $v_2$ for charged 
 particles in the centrality $(15-25)$\% Au+Au at 200A GeV, 
 computed with fluctuating IC. The vertical bars indicate 
 dispersions. The data are from PHOBOS Collab.\cite{phobos3}. 
 Right: $v_2$ distribution in the interval $0.48<\eta<0.95\,$, 
 corresponding to CP EoS and CE.} 
\label{fig:v2}
\vspace*{-.5cm}
\end{figure}
\medskip

\noindent{\bf Pseudo-rapidity distribution}: 
Figure \ref{fig:EoS2} shows that the inclusion of a critical 
end point increases the entropy per energy. This means that, given the same total energy, CP EoS produces larger 
multiplicity, which is clearly shown in the left panel of 
Figure \ref{fig:eta}, especially in the mid-rapidity region. 
Now, we shall mention that, once the equations of state are chosen, fluctuating IC produce smaller multiplicity, for the same decoupling prescription, as compared with the case of 
smooth averaged IC~\cite{review}. 

\noindent{\bf Transverse-Momentum Distribution}: 
As discussed in Sec. \ref{ingredients}, since the pressure does 
not remain constant in the crossover region, we expect that the 
transverse acceleration is larger for CP EoS, as compared with 
1OPT EoS case. In effect, the right panel of Figure 
\ref{fig:eta} does show that $p_T$ distribution is flatter for 
CP EoS, but the difference is small. 
The freezeout temperature suggested by $\eta$ and $p_T$ 
distributions turned out to be $T_f\simeq135-140\,$MeV. 
\medskip


\noindent{\bf Elliptic-Flow Parameter $v_2$}: We show, in 
Figure \ref{fig:v2}, results for the $\eta$ 
distribution of $v_2$ for Au+Au collisions at 200A GeV. As 
seen, CP EoS gives larger $v_2\,$, as a consequence of larger 
acceleration in this case as discussed in 
Sec.\ref{ingredients}. 
Notice that CE makes the curves narrower, as a consequence of 
earlier emission of particles, so with smaller acceleration, 
at large-$\vert\eta\vert$ regions. Due to the IC fluctuations, 
the resulting fluctuations of $v_2$ are large, as seen in 
Figures \ref{fig:v2}. It would be nice to measure such a 
$v_2$ distribution, which would discriminate among several 
microscopic models for the initial stage of nuclear 
collisions. 
\medskip 


\begin{figure}[!t]
\includegraphics*[scale=0.28]{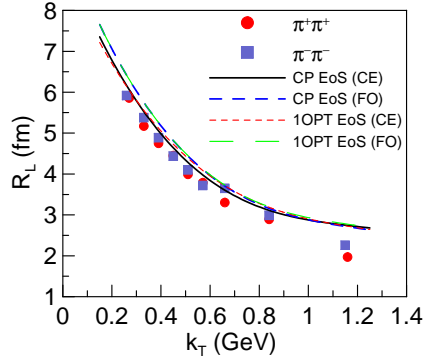}
\vspace*{-.5cm} 
\caption{$k_T$ dependence of HBT radius $R_L$ for $\pi$  
 in the most central Au+Au at 200A GeV, computed with   
 fluctuating IC. The data are from PHENIX Collab.\cite{phenix}.}
\label{fig:RL}
\end{figure}
\begin{figure}[!b]
\includegraphics*[scale=0.28]{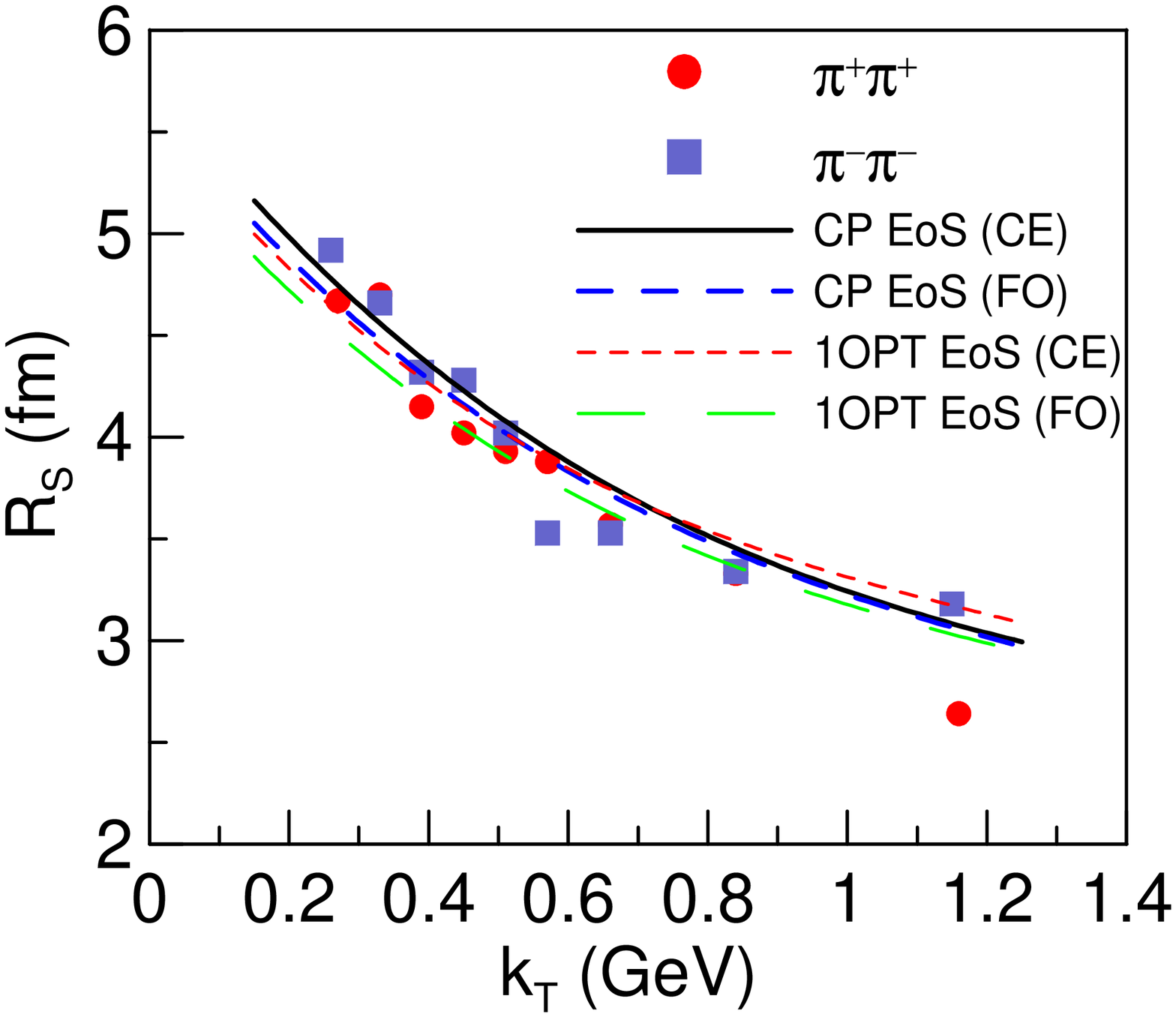}
\vspace*{-.5cm}
\vspace*{.cm}
\includegraphics*[scale=0.28]{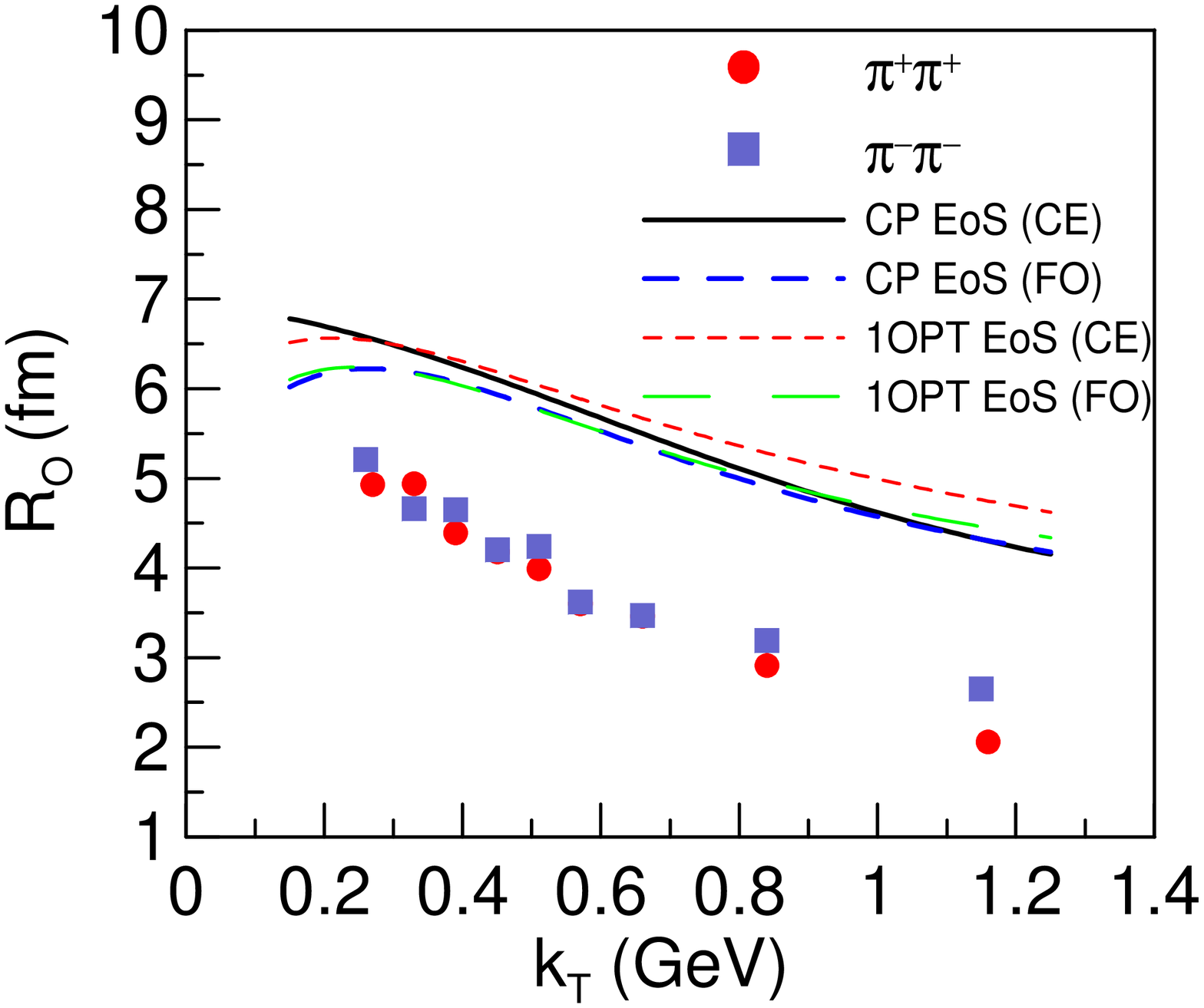}
\vspace*{-1.cm}
\caption{$k_T$ dependence of HBT radii $R_s$ and $R_o$ for 
 pions in the most central Au+Au at 200A GeV, computed with 
 event-by-event fluctuating IC. The data are from PHENIX 
 Collab.\cite{phenix}.}
\label{fig:RoRs}
\end{figure}
\noindent {\bf HBT Radii}: 
Here, we show our results for the HBT radii, in Gaussian 
approximation as often used, for the most central Au+Au 
collisions at 200A GeV. As seen in Figures \ref{fig:RL} and 
\ref{fig:RoRs}, the differences between CP EoS results and 
those for 1OPT EoS are small. 
For $R_s\,$, and especially for $R_o\,$, one sees that CP EoS 
combined with continuous emission gives steeper $k_T$ 
dependence, closer to the data. 
However, there is still numerical discrepancy in this case. 

\section{CONCLUSIONS AND OUTLOOKS}
\label{conclusions}  
In this work, we introduced a parametrization of lattice-QCD 
EoS, with a first-order phase transition at large $\mu_b$ and 
a crossover behavior at smaller $\mu_b\,$. 
By solving the hydrodynamic equations, we studied 
the effects of such EoS and the continuous emission. 
Some conclusions are: 
{\it i}) The multiplicity increases for these EoS in the 
 mid-rapidity;  
{\it ii}) The $p_T$ distribution becomes flatter, although the 
 difference is small;  
{\it iii}) $v_2\,$ increases; CE makes 
 the $\eta$ distribution narrower; 
{\it iv}) HBT radii slightly closer to data. 
   
In our calculations, the effect of the continuous emission on 
the interacting component has not been taken into account. A more realistic treatment of this effect probably makes $R_o$ 
smaller, since the duration for particle emission becomes 
smaller in this case. Another improvement we should make is 
the approximations we used for ${\cal P}(x,p)$.


\begin{theacknowledgments}
We acknowledge financial support by FAPESP (04/10619-9, 
04/15560-2, 04/13309-0), CAPES/PROBRAL, CNPq, FAPERJ and PRONEX. 
\end{theacknowledgments}

\end{document}